# Transmission laser welding of similar and dissimilar semiconductor materials

*Pol Sopeña, Andong Wang, Alexandros Mouskeftaras, and David Grojo\**

P. Sopeña, A. Wang, A. Mouskeftaras, D. Grojo
Aix-Marseille Université, CNRS, LP3, UMR7341, 13009 Marseille, France
E-mail: david.grojo@univ-amu.fr

Keywords: laser welding, laser processing, wafer bonding, semiconductors, silicon.

Laser micro-welding is an advanced manufacturing method today applied in various domains. However, important physical limitations have prevented so far to demonstrate its applicability in silicon and other technology-essential semiconductors. Concentrating on circumventing the optical limits on the deliverable energy density at interfaces between narrow-gap materials with intense infrared light, we make the first feasibility demonstration of transmission laser welding between silicon workpieces using nanosecond laser pulses. We obtain a shear joining strength of 32±10 MPa which compares very favorably to the complex process alternatives. Supported by experiments repeated on different material combinations including gallium arsenide, we confirm that this remarkable performance level is achievable for similar and dissimilar semiconductors. The demonstrations rely on a small footprint fiber laser, an aspect that holds great promises for the advent of a high-efficiency flexible process beneficial for important technology developments including lab-on-a-chip and hybrid semiconductor systems.

# 1. Introduction

Laser welding is today a key process in modern manufacturing. In this view, the use of tightly-focused ultrashort pulses has been an important breakthrough by providing the ability to achieve energy deposition by nonlinear absorption anywhere in the three-dimensional space inside transparent materials. Subsequent highly-localized material melting is the basis of femtosecond laser welding.[1] Various successfully demonstrated applications include through-glass,[1-3] through-polymer,[4] or through-ceramic configurations.[5] While micro-bonding would surely find immediate applications in the sector of microelectronics, it is however striking to realize that such process is not directly applicable to bond together different semiconductor workpieces. This problem is today addressed by the introduction of an absorbing layer (e.g. metal),[6] or even adhesives leading to solutions incompatible with the most demanding applications.

For alternative methods directly applicable to silicon (Si) and other semiconductors, one can refer to wafer molecular bonding. It consists in appropriately preparing and placing two wafers in the most intimate contact so that intermolecular bonds appear across the wafer.[7] However, the resulting bonding strengths remain relatively modest (few kPa) until the bonding is enhanced by processes like thermal annealing, plasma activation of the surface, or the application of electric fields. This yields typical bond strengths in the order of several MPa and a level of performance appropriate for some highly demanding applications.[8,9] However, a major drawback of the technique is the requirement of numerous and tedious steps in clean room environment for assembling functional devices. In this context, a more direct technology as laser micro-welding remains highly desirable for increased flexibility and the fabrication of complex-architecture semiconductor systems inaccessible by current methods.

In the race for a semiconductor laser welding technology, a critical step has been made very recently by translating the ultrafast regime into the infrared (IR) domain of the spectrum for a demonstration of Si-metal welding.[10] A major limitation in such a through-Si configuration relies on the strong propagation nonlinearities inherent to narrow gap materials which tend to defocus and delocalize intense IR radiation inside materials. This renders internal or through-semiconductor processing a very challenging task unless compensation measures are taken by adjusting the spatiotemporal characteristics of irradiation.[11] Among the available solutions, we can refer to non-conventional hyper-focusing conditions,[12] ultrafast trains of pulses to rely on accumulation processes,[13] or picosecond pulses for reduced power.[14] The later approach was the one applied for Si-metal welding.[10] Adding an advanced procedure for compensation of the remaining nonlinear focus shift with 10-ps pulses, precise interaction at a Si-copper interface could be achieved for welding. However, the obtained bonding strengths remained modest in comparison to the MPa bonding strengths routinely achieved in glass-metal welding studies.[15] As we will see later, besides the nonlinear propagation issues, this indicates other limitations related to the semiconductor interface problem. Given the favorable opacity given by a metal for the bottom material in that studied case, one can directly anticipate an even harder welding situation in a semiconductor-semiconductor configuration. While we do not expand on this in this manuscript, we confirm this hypothesis having made numerous unsuccessful attempts of Si-Si welding by using pulses at 1550-nm wavelength and pulse durations from 200 fs up to $\cong$ 20 ps (not shown here).

To circumvent the impossibility of directly translating the excellent performances of ultrafast laser glass welding to semiconductor welding, we propose in this work a disruption in the approach and the use of the long pulse regime. This is inspired by the *stealth dicing* technology, a major successful application of sub-surface modifications in semiconductors by nanosecond lasers.[16-18] The key aspect of the approach is to exploit a local thermal runaway accessible with near-IR nanosecond pulses and causing a locally melted volume near focus under the surface. This is the exact analog of the desired situation for welding except that the sub-surface modifications resulting from material re-solidification are taken as created weaknesses in a bare semiconductor for subsequent dicing. In welding applications, the same local melting would be intended at an interface to create bonds between two semiconductors. Interestingly, the emergence of erbium-doped nanosecond lasers emitting at 1550 nm, as the one used in this work, has motivated recent successful studies in the two-photon absorption regime to initiate similar localized thermal runaway.[17] As two-photon absorption depends on the square of the intensity, the main advantage is then the capacity it provides to deposit the laser energy at arbitrary depths inside Si. This leads to a process governed by thermo-mechanical effects resulting in novel 3D microprocessing solutions for creating complex functional structures as for instance buried microfluidic channels or waveguides in Si chips.[19-21]

In this work, we study the potential of this heat-driven processing method to solve the challenging problem of laser welding between Si and/or other semiconductors workpieces. The nanosecond laser regime leads to a situation that we extricate from the important nonlinear propagation features found in previous attempts with ultrashort pulses.[10] Precise localization of the energy deposition at the semiconductor interfaces becomes then achievable by simple pre-optimization procedures. In this way, we can concentrate on other limiting factors. We identify that the high refractive index inherent to narrow gap materials leads to a sensitivity to contact imperfections which has no equivalent in any other laser welding applications. Relying on regions in optical contact, we demonstrate for the first time the feasibility of Si-Si welding and successfully extend this work to gallium arsenide (GaAs). Supported by the measurement of shear joining strengths >10 MPa for all processed configurations, we reveal the potential of the approach for micro-welding of similar and dissimilar semiconductors. Finally, we show that centimetric areas can be readily processed with a very compact fiber-laser source and confirm the expected weld resistance to high-load in this case. This indicates the possibility for scaling up the process to large wafers with the use of industrial high-power sources and thus providing a unique semiconductor surface bonding solution suppressing some stringent preparation procedures.

## 2. Results and Discussion

### 2.1 The Prerequisite of an Optical Contact

To demonstrate the feasibility and reveal the critical parameters for laser welding of Si, we irradiate in this work Si-Si interfaces using an erbium-doped fiber source emitting pulses of <5 ns duration at 1550-nm wavelength; that is in the full-transparency domain of Si. Capitalizing on the mentioned previous works demonstrating the possibility of local energy deposition inside Si with tight focusing conditions,[14] but also potentially severe nonlinear focal shifts requiring pre-compensation for processed zone positioning,[11] we concentrate first on z-scanning procedures. The experimental approach is schematically represented in **Figure 1a**. In brief, it consists in repeating single-spot irradiations with 1000 applied pulses at the maximum available energy (11 µJ on target) and repetition rate (1 kHz) of our compact nanosecond laser with 0.45 NA (numerical aperture) focusing conditions. In this way, we produce well-contrasted modifications which can be observed *in-situ* by IR transmission microcopy. In Figure 1b-c, we can visualize the vicinity of the interface between the two wafers from a lateral view. The laser-induced modifications taking the form of black elongated marks (first modification indicated by a downwards triangle) and separated laterally are obtained by displacing the beam from top to bottom across the interface with depth change steps of ~12.5 µm between irradiations. In Figure 1b, as the beam focus is z-scanned downwards (dashed purple arrow), we observe how the produced modification is replicated crossing the interface (red dotted line) and finally gets fully confined inside the lower wafer. By separating the wafers and observing by visible microscopy the top and bottom surfaces in contact (Figure 1d), we identify circular features on both substrates taken as an encouraging observation as it shows the possibly to process together materials at both sides of the interface.

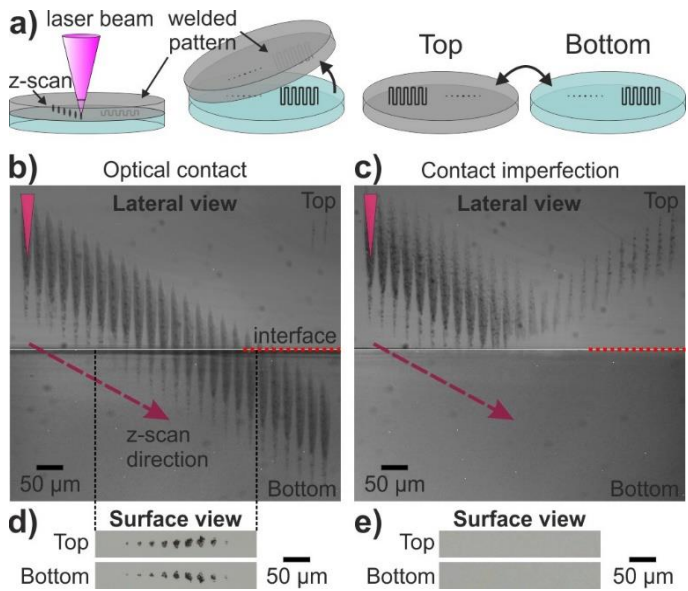

**Figure 1. a)** Schematic of the welding configurations (z-scan and pattern) and the consecutive separation procedure to visualize the modifications at surfaces of upper and lower wafers in contact during laser irradiation. **b-c)** Lateral view by IR transmission microscopy of the interface region of two Si wafers. The red dotted line indicates the interface placement. In the imaged zones, single spot modifications of 1000 pulses at 11 µJ are produced (25 µm lateral separations) at various depth (12.5 µm steps) from top to bottom (purple dashed arrow). **d-e)** Corresponding visible-light microscopy images of the surfaces of top and bottom wafers after sample separation as indicated in a). **b,d)** Processed zone with modifications under the interface indirectly evidencing a good optical and efficient coupling to the bottom wafer. **c,e)** Processed zone with modifications inside the upper sample with beams reflected by the interface and absence of surface modification due to imperfect contact.

Repeating this process in different parts of the sample, we realize it is not always reproducible due to non-uniformity of the contact. This is not surprising as any laser welding study would show that the materials should ideally be placed in intimate contact during processing to obtain strong and reliable bonds. In practice, it is admitted that a material mixture with gaps larger than few micrometers is hardly achievable.[22,23] An advantage for achieving Si welding is the high specifications for microelectronics grade wafers with sub-nanometer roughness and excellent flatness. This is favorable to deal with this nontrivial mechanical problem but a complication originates from its high refractive index (n=3.5) making any contact imperfection a resonating optical cavity. The existence of small air gaps between the top and bottom polished wafers can cause a nearly total reflection as evidenced on Figure 1c by the well-defined modifications visible in the upper sample after processing with beams tentatively focused inside the lower wafer. Considering this optical problem, it is only when the gap is considerably smaller than the wavelength that an optical contact is achieved and the beam can easily propagate through the interface without attenuation (Figure 1b). For larger gaps, the interface acts as a Fabry-Perot interferometer and the transmission to the bottom wafer becomes limited (Figure 1c). Given the high refractive index of Si and its associated high reflectance (R~30%), one predicts that this becomes an extremely more critical aspect than for previous studies in glass or polymer materials. These optical considerations are confirmed with Figure 1e showing a total absence of modification at the surface of the bottom substrate as a consequence of a modest local transmission by the Fabry-Perrot cavity. A more surprising feature is the total absence of visible modification also on the top wafer surface while the lateral view reveals clearly that some internal modifications of the top sample are sectioned by the interface. In comparison to Figure 1b and 1d, this indicates a significantly reduced field on both surfaces and so obviously inappropriate conditions for welding.

A major conclusion from these first observations is the imperative requirement of a nearly perfect optical contact to tentatively achieve welding of semiconductors or other high index materials. Obviously, it is a nontrivial problem that scales with the required contact area. For the following investigations we rely on high-grade polished wafers of typically 15×18 mm$^2$ surface-area and identical preparation procedures are systematically repeated including cleaning steps and the application of a mechanical clamping pressure on the contacting samples (see details in Section 4). As we will see later on, despite this procedure, slight local variations of the interface gap remain, as observed indirectly through the beam processing capability. However, these non-uniformities have been substantially diminished after carefully elaborated tests and sample preparation optimizations. In this report concentrating on Si and GaAs, these technical developments have been crucial for reproducible results and reliable statistical analyses of welding performances.

**2.2 Evidence of Material Mixture in Line-writing Configuration**

As a next step towards welding, we produce continuous line-shaped modifications with repeated pulses employing the same irradiation conditions (0.45 NA, 11 µJ, 1 kHz). As for the above z-scanning approach with static irradiations and lateral IR imaging, we test the response to different focal shifts with respect to the interface. Each line is 100 µm long and is obtained after a relative motion of the beam parallel to the interface plane at a speed of 2 µm s$^{-1}$, corresponding to ~7500 applied pulses according to the 1 kHz repetition rate of our laser and ~15 µm modification size as estimated from Figure 1d. In **Figure 2a**, we observe optical microscope images of both samples after separation. This leads to a mirror symmetry between top and bottom sample observations as schematically represented in Figure 1a. We define as a reference for comparison the position $z_0$ corresponding to the most centered modification on the interface (based on Figure 1b) and we vary the focusing distance in steps of 12.5 µm inside Si above (positive) and below (negative) the interface (the lateral view IR image is shown in Supporting Information). Upon surface inspection, we observe first thin lines with widths of ~7 µm when the beam is focused approaching from above the interface. These become wider (~20 µm) and darker as we move towards the interface and then start fading away when the beam is focused inside the lower wafer. This transition corresponds to a cross-sectional mapping of the beam in terms of spatial energy distribution, as it is suggested from the lateral view modifications in Figure 1b. Comparing the top and bottom wafers we can observe a clear one-to-one mapping of the contour of the lines, considering one is the mirror image of the other. This already indicates a modification regime supported by both surfaces which is a favorable observation but does not demonstrate material mixture.

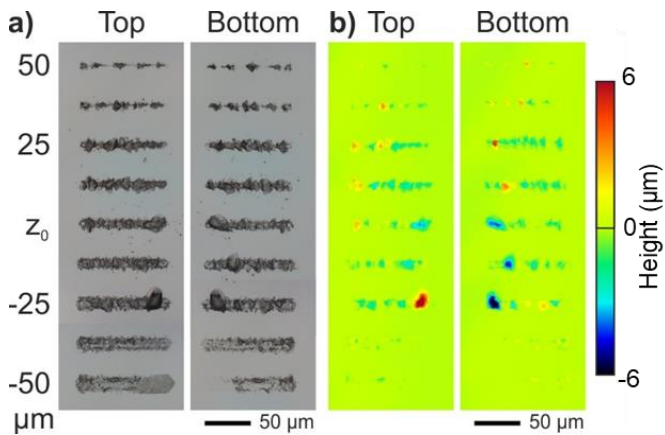

**Figure 2. a)** Top and bottom wafer surface optical microscopy images of the produced lines at the interface at different focal distances with respect to reference position $z_0$ (see Supporting Information). **b)** Corresponding confocal microscopy images revealing the surface topography.

Both the contour and topography of the top and bottom lines match each other indicating material transfer between wafers. Note the mirror-like correspondence of the processed zones between top and bottom images as represented in Figure 1a.

Material transfer between samples becomes evident when we analyze the surface topography with the corresponding confocal microscopy images shown in Figure 2b. From these images, we observe that not only the line contours but also the topography profiles on both wafers match very well, one being the negative version of the other. The micrometer-dimension features can be reasonably taken as an evidence of material exchanges in the processed zones and thus possible welding, finding the most material transfer at focusing conditions around the reference position ($z_0$). This already represents a very promising result showing clearly the feasibility of Si-Si micro-welding from a simple laser configuration but one could not prejudge the Si bonding performance from it before its evaluation by shear joining strength measurements.

**2.3 Bond Strengths**

For further optimizing the irradiation conditions and perform measurements on the highest achievable bond strengths in our configuration, we investigated the influence of the number of applied pulses. For that, we produced 200 µm lines at different scan speeds (from 1 µm s$^{-1}$ to 5 mm s$^{-1}$) with the same irradiation conditions as before (0.45 NA, 11 µJ, 1 kHz). A representative selection of the optical images of the bottom wafer lines is shown in **Figure 3a** for the Si-Si configuration. A more complete set of observation including also the inspection of top wafer surfaces for this and other tested material combinations (described hereafter) is given in Supporting Information. As one could expect, we observe on the Si-Si case that the widest (~15 µm) and most pronounced marks and so the largest processed volumes exhibiting material mixture, are obtained at the lowest scanning speed, corresponding to the highest number of applied pulses and so the maximum of incubation benefit. Given the need to process significantly larger areas for measurable bond strengths, we have chosen as a compromise between line quality and scan speed, 2 µm s$^{-1}$. Obviously, these optimum scanning conditions are directly dependent on the specifications of the laser used leaving room for further optimization in future studies with the use of sources delivering higher powers and/or repetition rates.

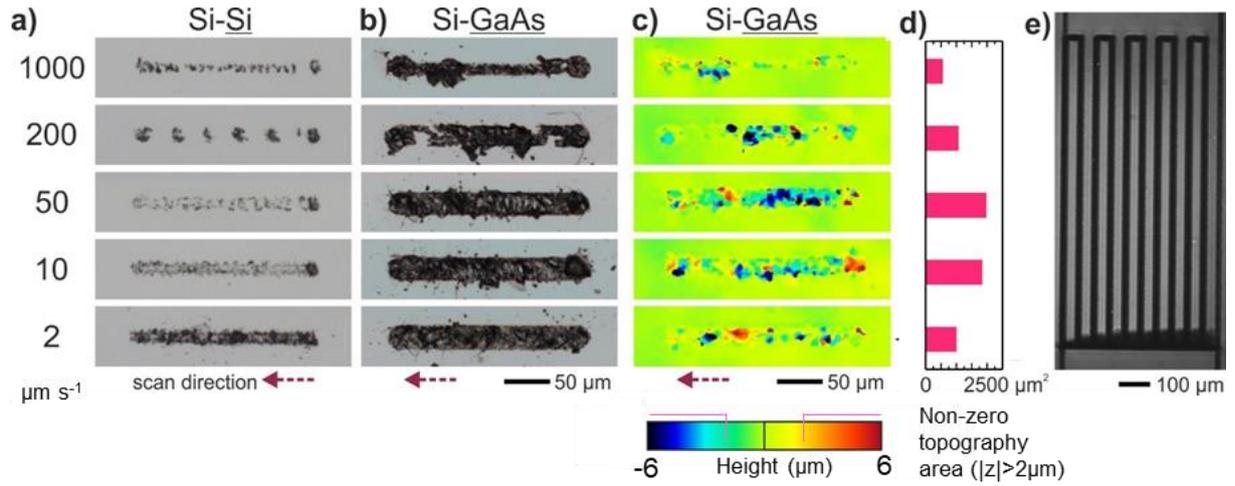

**Figure 3.** Optical microscopy images of the welded lines on the bottom wafer for the **a)** Si-Si and **b)** Si-GaAs configuration. The best scanning conditions at 11 μJ correspond to 2 and 50 μm s$^{-1}$, respectively. **c)** Topography images by confocal microscopy corresponding to the images in b). **d)** Corresponding measured areas (c) exhibiting surface elevations or depressions exceeding a threshold fixed at 2 μm. **e)** IR transmission image of a welded serpentine in the Si-GaAs configuration. Based on estimations the total welded area of ~0.25 mm$^2$ should resist shear force up to 4.5 N.

Using these pre-optimized conditions (1550 nm, 11 μJ, 0.45 NA, 1 kHz, scanning speed 2 μm s$^{-1}$), we then repeat raster-scanning procedures similar the one shown in Figure 3e at the center of several samples for statistical evaluation of the weld strengths analyzed in terms of shear joining strength. A commercial mechanical force tester is used for this purpose. The details of the measurement procedure are given in Section 4. Briefly, we rely on a linearly increasing shear force applied on the welded sample while a loadcell is used to record the force levels for which samples are separated. According to the typical <0.5 mm$^2$ processed area (see Figure 3e), the 10-N measurement limitation of the used loadcell corresponds to a maximum measurable weld strength below the ultimate tensile strength of 7 GPa of monocrystalline Si.[24] This makes an appropriate experimental configuration for accurate measurements of welds at MPa levels. Repeating the experiments on four Si-Si samples, we conclude on a shear tensile strength of 32±10 MPa before weld fracture. This value is obtained by calculating the ratio between the measured force and the apparently welded area which is a function of contact imperfections on the different samples but can be determined by microcopy after the wafers are separated (see Section 4). This measurement becomes the first evidence of Si laser welding obtaining an excellent bonding strength in comparison to alternative non-laser bonding techniques (including adhesive or molecular wafer bonding).[8,9] The measured strength actually compares very favorably with the best ultrafast laser welding performances demonstrated on dielectrics for micro-optics applications.[3]

For technological considerations, one can expect a breakthrough, if similar performance level is achievable on other semiconductors but also on dissimilar semiconductor configurations. In that respect we add another semiconductor to our considerations: GaAs. Among other differences, GaAs differs to Si by its direct bandgap, which makes it a widely used material in the industry for different applications as important as light emitting diodes. Compared to Si, GaAs exhibits a similar refractive index (n~3.4) and reflectance (R~30%) but a higher two-photon absorption coefficient of 50 cm $GW^{-1}$ given its direct bandgap (1.3 cm $GW^{-1}$ for Si).[25,26] An important consequence is a drastically reduced laser fluence threshold for modification.[27] Following a similar strategy as before, we first substitute the Si bottom wafer for a GaAs one and then vary the scan speed at constant pulse energy (11 μJ) to determine the optimum irradiation conditions. In Figure 3b we show a representative selection of images of the lines produced on the bottom GaAs. The lower modification threshold of GaAs allows stronger interactions resulting in wider (~30 μm) and more visible lines. However, unlike before, analyses of the confocal images (Figure 3c) reveal an optimal scan speed that corresponds to 50 μm $s^{-1}$ instead of the slowest one. To illustrate the detailed analyses on this aspect (see also Supporting Information), we present in Figure 3d the areas exhibiting elevations or depressions in the topography exceeding 2 μm (|z|>2 μm) for each image of Figure 3c. By taking this arbitrary criterion as an evaluation of the probability for material mixture, we conclude on an optimum scan speed of 50 μm $s^{-1}$. This increased speed in comparison to Si allows faster raster-scanning irradiations on several samples (shown in Figure 3e) leading to a measured shear joining strength of 18±1 MPa for this dissimilar semiconductor configuration. The difference compared to Si-Si can be reasonably attributed to the usual challenges associated with the welding of materials exhibiting different thermo-mechanical properties.[28] However, it is modest and the measured strengths show interestingly the potential of this technique for welding Si with other semiconductors using a through-Si transmission configuration.

For an even more general demonstration, we repeat the studies with through-GaAs configurations. GaAs samples are then placed on top of both Si and GaAs and pre-optimization studies similar to the previous cases are conducted. The optical microcopy images obtained to determine the best processing conditions are shown in Supporting Information. In this case, an important change is to decrease the pulse energy down to 1.4 µJ to achieve localized energy deposition and well-defined internal modifications. At the maximum laser energy used for through-Si processing (described above), the higher propagation nonlinearities in GaAs, including nonlinear absorption, are actually developing far prior to focus preventing enough energy delivery near focus to achieve modifications.[25,29] For both configurations, the best scanning speed is 20 µm s$^{-1}$. In these cases, the shear joining strength is measured at 22 and 16±6 MPa for GaAs-GaAs and GaAs-Si, respectively. This definitively confirms the feasibility of nanosecond laser welding applied to semiconductors and achievable bond strengths >10 MPa for similar and dissimilar materials when employing adequate irradiation conditions.

Comparing the obtained values of shear joining strengths in **Figure 4**, it is striking to note that these are all in the same order of magnitude, with Si-Si exhibiting the strongest bond. Dissimilar material welding configurations, with either Si or GaAs on top, result in a similar value, both smaller than those obtained with Si-Si. Despite a single measurement on the GaAs-GaAs welding case, the result indicates a performance at the level of the Si-Si samples (bottom of measurement distribution) and thus is consistent with a superior performance achieved for similar semiconductor welding. In parallel, it is however interesting to note the reduced statistical dispersion of measurements in the Si-GaAs configuration. This indicates a better robustness to experimental fluctuations, an aspect that can be also attributed to the significantly lower modification threshold of GaAs compared to Si.[27] With this consideration, we expect that the Si-GaAs configuration must tolerate more the imperfections of the optical contact as, even at the expected interface transmission of about 50% for two well separated surfaces (R=30%), material melting can be readily reached together in Si and GaAs lower sample.

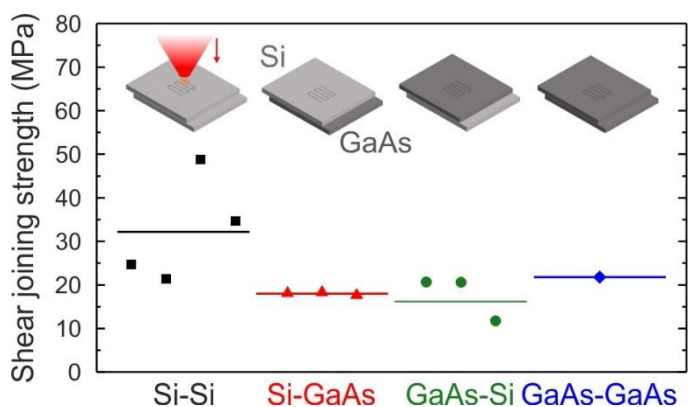

**Figure 4.** Joining strengths as evaluated from shear force measurements (markers) before weld fracture for all the different configurations (indicated below). The average value of each configuration is represented with a straight continuous line.

Comparing now the obtained shear forces with those reported for state-of-the art wafer bonding technologies (~15 MPa),[9] we obtain comparable and even slightly better results but, more importantly, we introduce in this context some advantages of simplicity and flexibility. For instance, the methodology used in this work implies much less surface preparation efforts in comparison to the tedious and multiple steps performed in clean room environment for the best wafer bonding methods (e.g. molecular anodic bonding). However, an even more important advantage on the laser approach is on the capacity to rely on a local and digital technique. This makes possible to join small pieces without compromising a whole wafer. It gives also access to complex architectures (e.g. curved surfaces) and multi-semiconductor configurations that would not be possible otherwise. This must open direct perspectives for new practices in the manufacturing of microelectromechanical systems (MEMS) and an interesting process for emerging new concepts based on hybrid chips. The latter includes for instance co-designed electronics and microfluidics systems for sustainable cooling solutions.[30,31] In this comparison to wafer bonding a drawback remains with the affected processed layer of few tens of micrometers as shown with the lateral views in Figure 1b-c. This compromised interface zone is inherent to laser techniques as the results obtained here on semiconductors compare very well with previous works accomplished in very different materials (e.g. dielectrics, polymers, ceramics).[2-5,28] Obviously, this issue will not be solved without combining the best of each technique, that is preparing a molecular contact similar to the wafer bonding method before achieving a local material melting and perfect recrystallization by appropriate laser irradiation. While this was obviously out of scope for the present paper aiming at demonstrating the general feasibility of semiconductor laser welding, it supports a vision for possible process improvements.

## 3. Conclusion

This work proves the feasibility of laser welding of semiconductors using a very compact nanosecond laser technology and relatively loose focusing. In that regard, it must introduce a solution with great application potential. To our knowledge, it provides not only the first demonstration of laser welding for similar semiconductors but also dissimilar ones (Si and GaAs). The achieved bonding strengths in all configurations are on the order of tens of MPa, similar to those obtained by wafer bonding techniques and ultrafast laser welding of transparent dielectrics, a technique inapplicable for semiconductors with conventional laser processing configuration. In the context of these comparisons, the obtained results lead to unprecedented process efficiency. To illustrate this aspect, we realized that a weld of only few millimeters with the reported strengths must allow hanging from it the laser equipment that we employ to achieve welding. For this final evidence of performance shown in **Figure 5** we use the Si-GaAs configuration as it allows higher processing speeds. Given the remaining sensitivity to contact non-uniformities, a problem scaling with the intended contact area, we intentionally processed a relatively large zone. With a total processed area of about 3.5×7.0 mm$^2$ shown by IR imaging in Figure 5b, the shear joining strength of 18 MPa measured in this configuration (see Figure 4) translates in a resistance to a shear force up to 400 N. This is one order of magnitude above the applied shear force of about 30 N resulting from hanging on the laser head (~3 kg). These considerations explain the successful demonstration shown in Figure 5a despite imperfect contact as observed in the IR transmission image of Figure 5b.

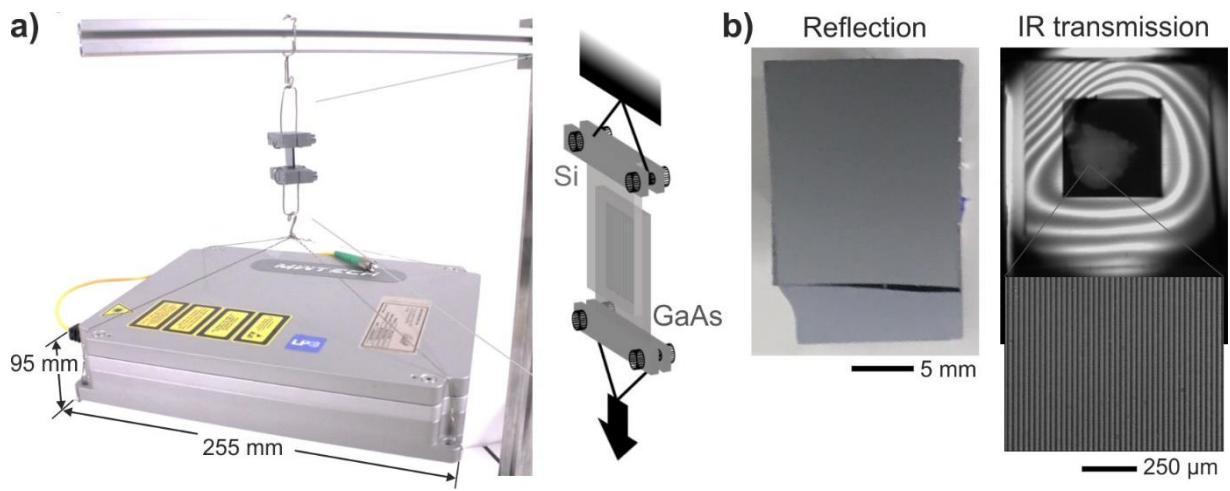

**Figure 5. a)** Image of the laser source used in the experiments (~3 kg) hanging from a Si wafer welded on GaAs. A schematic representation is shown at the right. b) From the IR transmission image, we observe the total processed area of 3.5×7.0 mm$^2$ from which we estimate a resistance to a shear force up to 400 N. The interference fringes visible on the IR image are attributed to the Fabry-Perot cavity at the material interface and then reveal the contact non-uniformity.

To expand on the most appropriate laser interaction regime for semiconductor applications, one should first emphasize on the tremendous trend towards the use of ultrafast laser technologies for transparent material welding. This is today fully justified for highly localized and controllable energy deposition in/trough dielectrics. The successful glass-glass or glass-Si ultrafast laser welding demonstrations with reported shear joining strength exceeding 50 MPa for some cases are clearly supporting a superiority to nanosecond laser welding in this context.[22,32,33] However, it is worth highlighting here that similar ultrashort pulses in the IR domain simply fail in semiconductors. This is an aspect we confirmed experimentally (not shown) and which is consistent with the recent literature reporting the strong delocalization of intense IR light inside semiconductors due to strong propagation nonlinearities.[11] This statement is also in line with the recent successful demonstration of Si-copper with picosecond laser pulses reporting, despite advanced nonlinearity compensation measures, limited weld strengths (maximum shear bonding strengths up to ~2 MPa) in comparisons to through-glass ultrafast laser welding configurations widely investigated during the last decades.[10]

Accordingly, our advance is somehow based on a step back returning to the longer pulse regimes originally used in the first attempts of laser welding on different materials. By taking a direction that may appear counter-intuitive, this work solves the long-standing problem of semiconductor welding. Compared to current wafer bonding and the laser micro-welding methods applied on other materials, our method represents a very cost-effective and flexible solution as it allows dissimilar material assembly and does not require clean room environment. Based on our demonstration, we anticipate various optimizations will derive rapidly depending on targeted applications. For instance, using high-powered industrial sources, the process will be easily scaled up for increased process efficiency. We extrapolate directly from our methods that scan speeds of several m s$^{-1}$ should be accessible with MHz repetition rates and/or milli-Joule energy levels. All in all, we highly expect that the identified nanosecond laser solution for micro-welding of similar and dissimilar semiconductors opens the door for new high-value manufacturing practices in the semiconductor industry.

## 4. Experimental Section

*Laser system:* The laser experiments are carried out using an erbium-doped fiber laser source (MWTech, PF1550) delivering <5 ns (FWHM) pulses at 1-kHz repetition rate and 1550-nm central wavelength. The housing of the small footprint laser head can be seen in Figure 5a. After the laser, a telescope (focal length of 100 and 200 mm) is used to expand the beam and minimize the residual beam divergence from the fixed-focus commercial collimator (Thorlabs KAD12NT) installed at the laser fiber output. To control the delivered pulse energy, a half-wave plate and a polarizer combination is used in the beam path before the focusing optics. Accounting for the transmission of all optical elements in the beam path the maximum delivered energy with focused beam is ~11 µJ per pulse on target as measured by employing a thermopile-based laser power sensor (Ophir, 3A-P-V1).

*Samples and assemblies for welding:* We employ two different semiconductors, Si and GaAs. Si samples are pieces of 15×18 mm$^2$ which are individually cut from 4-inch wafers of 1-mm thickness (Siltronix, orientation (100)+/-0.5°, FZ growth method, intrinsic, resistivity >900 Ω·cm,). For the GaAs sample same-size pieces are cut from 4-inch diameter wafers of 600-µm thickness (Neyco, orientation (100)+/-0.3°, VGF growth method, intrinsic). All initial wafers are double-side polished ensuring a surface roughness <1 nm (RMS). The prepared samples are thoroughly cleaned to remove any organic material by wiping it with optical paper soaked in acetone and later rinsed with dry air. To determine appropriate cleaning procedure, we initially cleaned the wafers in an ultrasound bath and atmospheric pressure plasma (ULS Omega, AcXys Technologies) but we observed no difference in the final optical contact quality so we later omitted these steps. Finally, the samples are stacked on a custom holder with supporting screws which are tightened to ensure a uniform clamping force during processing and improve the wafer contact at the interface.

*Welding experiments:* During the laser-semiconductor interaction experiments, the holder is mounted on motorized stages allowing fine three-dimensional positioning of the sample interface with respect to the beam focus. Absolute repositioning of the focusing depth is guaranteed by an *in situ* microscopy arrangement taking the observation of the front surface as a reference. For focusing the beam on the interface below the surface, we employ IR microscope objectives equipped with a correction collar for spherical aberration pre-compensation (LCPLN-IR, Olympus). Different NA of 0.85, 0.65, and 0.45 were originally tested, all permitting localized internal modifications near interface with our laser. Considering that long working distance and Rayleigh length are beneficial to avoid positioning complications, we have decided to concentrated the investigations on the 0.45 NA case for the feasibility demonstrations targeted in this work.

*Analyses of the welds:* For internal inspections of the welds, we use a custom IR transmission microscopy arrangement consisting of an IR InGaAs array (Raptor, OWLSWIR 640), a tube lens, and long working distance microscope objectives (Mitutoyo Plan Apo NIR series). For non-coherent uniform illumination, we use a quartz-tungsten halogen lamp. The same IR imaging system is used for lateral and top view imaging of the modifications created at the interfaces between stacked semiconductors. After the separation of samples, the top and bottom samples are also inspected by visible light reflection microcopy (Nikon Lv-UEPI-N) so that we can differentiate the effects on surfaces separated from internal modifications. Additionally, to evidence material transfer between samples, confocal microscopy (Leica, DCM 3D) is used for surface profilometry with nanometric longitudinal resolution. A commercial software (Leica, LeicaMap) is used for the image data analyses. The welding strengths are measured using a commercial mechanical force tester (LF-Plus Lloyd Instruments). The system is equipped with a loadcell (Lloyd Instruments, XLC Series) appropriate for load measurements up to 10 N. The measurement procedure relies on a shear force increasing at a linear rate of 0.05 mm min$^{-1}$ until the stack of semiconductor samples disjoins. The shear joining strength is obtained by calculating the ratio between the required force for sample separation and the welded area evaluated by optical microscopy of the surfaces after sample separation. While we based our shear strength measurements on the continuous highly-contrasted modifications observed by optical microcopy (see Figure 2a and 3a-b for example), we could have considered only the surfaces of non-zero topography revealed by confocal microscopy. In that case, the welded area would be smaller and would depend on the arbitrary topography threshold. On the basis of these considerations, the excellent welding performances reported in this work are actually conservative.

**Acknowledgements**

This work has received funding from the European Research Council (ERC) under the European Union's Horizon 2020 research and innovation program (grant agreement No 724480)